\documentclass[prl,groupedaddress,reprint,superscriptaddress,floats,nofootinbib,preprintnumbers]{revtex4-1}
\usepackage{graphicx} 
\usepackage{float}
\usepackage{amsmath}
\usepackage{epstopdf}
\usepackage{xspace}
\usepackage{rotating}
\usepackage{longtable}
\usepackage{multirow}
\usepackage{booktabs}
\usepackage[normalem]{ulem} 
\usepackage[breaklinks=true]{hyperref}
\hypersetup{
  colorlinks=true,
  linkcolor=blue,
  citecolor=blue,
  urlcolor=blue
}

\graphicspath{{./figs/}}
\usepackage{array,tabularx,epsfig,mathrsfs,rotating}
\usepackage{ifthen}
\usepackage{amsfonts}
\usepackage{amssymb}
\usepackage{slashed}

\renewcommand{\to}{\rightarrow}

\newcounter{diagram}

\newcolumntype{C}[1]{>{\centering\let\newline\\\arraybackslash\hspace{0pt}}m{#1}}
\begin{document}

\begin{flushleft}
IFT-UAM-CSIC-22-71
\end{flushleft}
%

\title{\boldmath
Baryogenesis from spontaneous CP violation in the early Universe
}
\author{S. J. Huber}
\email[]{S.Huber@sussex.ac.uk}
\affiliation{Department of Physics and Astronomy, University of Sussex, Brighton, BN1 9QH, UK
}

\author{K. Mimasu}
\email[]{ken.mimasu@kcl.ac.uk}
\affiliation{Department of Physics, King's College London, Strand, WC2R 2LS London, UK}

\author{J. M. No}
\email[]{josemiguel.no@uam.es}
\affiliation{Instituto de Fisica Teorica, IFT-UAM/CSIC,
Cantoblanco, 28049, Madrid, Spain}
\affiliation{Departamento de Fisica Teorica, Universidad Autonoma de Madrid,
Cantoblanco, 28049, Madrid, Spain
}

\hfill\draft{ }

\begin{abstract}
We propose a variant of electroweak-scale baryogenesis characterized by the spontaneous breaking of the charge-parity (CP) symmetry in the early Universe driven by the vacuum expectation value of a CP-odd scalar. This CP breaking period in the early Universe would be ended by the electroweak phase transition, with CP being (approximately) conserved at present, thus avoiding the stringent electric dipole moment experimental constraints on beyond-the-Standard-Model sources of CP violation. We study an explicit realization via a non-minimal Higgs sector consisting of two Higgs doublets and a singlet pseudoscalar (2HDM + $a$). 
We analyze the region of the 2HDM + $a$ parameter space where such an early Universe period of CP violation occurs, and show that the required thermal history and successful baryogenesis lead to a predictive scenario, testable by a combination of LHC searches and low-energy flavour measurements. 
\end{abstract}

\maketitle




\noindent \textbf{Introduction.}~Charge-Parity (CP) is an approximate symmetry of Nature, only broken in the Standard Model (SM) by  mixing in the fermion sector and the presence of three families of quarks and/or leptons~\cite{Cabibbo:1963yz,Kobayashi:1973fv}. The presence of 
CP violation in the early Universe, together with baryon number violation (which occurs at high temperature in the SM~\cite{Kuzmin:1985mm} via sphaleron processes) and a departure from thermal equilibrium, is required to generate the observed baryon asymmetry of the Universe (BAU)~\cite{Sakharov:1967dj,Cohen:1993nk,Trodden:1998ym,Morrissey:2012db,Konstandin:2013caa}. Although the amount of CP violation present in the SM is well-known to be insufficient for the generation of the BAU (baryogenesis) at the electroweak (EW) scale~\cite{Gavela:1993ts,Gavela:1994ds,Gavela:1994dt}, new sources of CP violation beyond the SM are tightly constrained by experimental searches of electric dipole moments (EDM) of the electron~\cite{ACME:2018yjb}, neutron~\cite{nEDM:2020crw} and atomic elements like mercury~\cite{Griffith:2009zz}. This constitutes a severe problem for the feasibility of scenarios for EW baryogenesis.

However, such tension between successful EW baryogenesis and current experimental EDM constraints could be circumvented by a period of spontaneous CP breaking in the early Universe, which would act as a catalyser for baryogenesis.~Then, the absence of new sources of CP violation beyond the SM at present times would naturally avoid the EDM limits.\footnote{See~\cite{Espinosa:2011eu,Cline:2012hg,Baldes:2016rqn,Cline:2017qpe,Carena:2018cjh,Hall:2019ank} for other weak-scale baryogenesis setups which avoid current  EDM experimental constraints via suppressed beyond the SM contributions to EDMs.} We show in this letter that 
this setup can be easily accommodated by a suitable extension of the SM Higgs sector. 
Furthermore, such a non-minimal Higgs sector can, at the same time, yield a strongly first-order EW phase transition (see e.g.~\cite{Cline:1996mga,Profumo:2007wc,Espinosa:2011ax,Dorsch:2013wja,Basler:2016obg,Dorsch:2017nza}), providing the departure from thermal equilibrium needed for baryogenesis.

Specifically, we consider a two-Higgs-doublet model (2HDM) with an additional $SU(2)_L$-singlet pseudoscalar $a$ (2HDM + $a$). This model has been considered recently as a well-motivated portal to dark matter (DM)~\cite{Ipek:2014gua,No:2015xqa,Goncalves:2016iyg,Bauer:2017ota,Abe:2018bpo,Robens:2021lov}, and we will discuss the resulting interplay between baryogenesis and DM in an upcoming work~\cite{HMN}, focusing here purely on achieving baryogenesis.   
In our setup, a vacuum expectation value (vev) of the pseudoscalar $a$ in the early Universe triggers the spontaneous breaking of CP. CP is then restored after the EW phase transition, which is first-order from the existence of a 
potential barrier between the CP violating (CPV) and EW minima. The region of parameter space that accommodates this thermal history of the Universe and leads to successful baryogenesis leaves no trace in current EDM experiments, but can be probed via current/future LHC searches and (possibly) via low-energy flavour experiments (rare $B$-meson decays). 
%
%


\vspace{2mm}



\noindent \textbf{2HDM $\boldsymbol{+}$ $\boldsymbol{a}$.}~Let us begin with the scalar potential for the model, $V = V_{\rm 2HDM} + V_{a}$. $V_{\rm 2HDM}$ is the 2HDM scalar potential for the two Higgs doublets $H_{1,2}$ with a softly-broken $\mathbb{Z}_2$ symmetry:
\begin{eqnarray}	
\label{2HDM_potential} 
V_{\rm 2HDM} &= &\mu^2_1 \left|H_1\right|^2 + \mu^2_2\left|H_2\right|^2 - \left[\mu_{12}^2 \, H_1^{\dagger}H_2+\mathrm{h.c.}\right] \nonumber \\
&+& \frac{\lambda_1}{2}\left|H_1\right|^4 +\frac{\lambda_2}{2}\left|H_2\right|^4  
+ \lambda_3 \left|H_1\right|^2\left|H_2\right|^2 \nonumber \\
&+& \lambda_4 \left|H_1^{\dagger}H_2\right|^2 + 
\frac{1}{2}\left[\lambda_5\left(H_1^{\dagger}H_2\right)^2+\mathrm{h.c.}\right]\, , 
\end{eqnarray}
and $V_a$ the potential involving the pseudoscalar singlet field $a$, 
\begin{eqnarray}
\label{Doublet_Singlet_potential} 
 V_{a} &=& \frac{\mu^2_{a}}{2}\,a^2 + \frac{\lambda_a}{4} a^4 + 
 \left(i\,\kappa\,a \,H_1^{\dagger}H_2 + \mathrm{h.c.}\right) \nonumber \\
 &+& \lambda_{aH_1}\, a^2 \left|H_1\right|^2 + \lambda_{aH_2}\, a^2 \left|H_2\right|^2
 \,.
\end{eqnarray}
CP conservation in the scalar sector at zero temperature imposes that the field $a$ have a vanishing vev $\left\langle a \right\rangle = 0$ in the EW vacuum, requiring 
\begin{equation}
\label{muS2_T0}
\mu^2_{a} + \left(\lambda_{aH_1}\, v_1^2  + \lambda_{aH_2} \, v_2^2 \right) = \mu^2_{a} + \lambda_{\beta} v^2 \equiv m_a^2 > 0,
\end{equation}
with $v_{1,2} = \sqrt{2} \left\langle H_{1,2}\right\rangle$ the vevs for the Higgs doublets after EW symmetry breaking (EWSB), $v = \sqrt{v_1^2 + v_2^2} = 246$ GeV the EW scale and $\lambda_{\beta} \equiv (\lambda_{aH_1} + \lambda_{aH_2}\, t_{\beta}^2)/(1+ t_{\beta}^2)$,  
%
%
with $ t_{\beta} \equiv {\rm tan} \beta = v_2/v_1$. 

\vspace{1mm}

In addition to the 125 GeV Higgs boson $h$, the physical scalar spectrum of the (CP-conserving) 2HDM contains a CP-even neutral state $H_0$, a CP-odd neutral state $A_0$ and a charged scalar $H^{\pm}$. Upon EWSB, the coupling $\kappa$ in $V_a$ induces a mixing between $a$ and $A_0$, giving rise to two mass eigenstates, which we denote by $a_{1,2}$ (with $m_{a_2} > m_{a_1}$). The  corresponding singlet-doublet mixing angle $\theta$ is given by (see e.g.~\cite{Ipek:2014gua,Goncalves:2016iyg,Bauer:2017ota}) $t_{2\theta} = 2\, \kappa \, v/(m_{A_0}^2 - m_a^2)$,
%
%
%
%
with $m_{A_0}$ the 2HDM $A_0$ mass. 
%
%
%
%
%
%
%
%
In the rest of this work we consider the 2HDM alignment limit~\cite{Gunion:2002zf} (favored by current LHC Higgs signal strength measurements~\cite{CMS:2020gsy,ATLAS:2020qdt}), in which an angle $\beta$ rotation of the CP-even neutral 2HDM field directions $h_{1,2}$ yields the CP-even Higgs mass eigenstates $h$ and $H_0$. In addition we 
fix for simplicity a common mass scale $M$ for the 2HDM states $H^{\pm}$, $H_0$ and $A_0$:
$m_{H_0}^2 = m_{A_0}^2 = m_{H^{\pm}}^2 = M^2 \equiv \mu_{12}^2/(s_{\beta}c_{\beta})$ (with $s_{\beta} \equiv {\rm sin} \beta$, $c_{\beta} \equiv {\rm cos} \beta$). 



%
%
%
%


\vspace{2mm}

\noindent \textbf{Early Universe Thermal History.}~The spontaneous breaking of CP in the early Universe requires a negative mass-squared term for the singlet field $a$ in~\eqref{Doublet_Singlet_potential},
\begin{equation}
\label{mu2_negative}
\mu_a^2 = s_{\theta}^2 \,m_{a_2}^2 + c^2_{\theta} \, m_{a_1}^2 - \lambda_{\beta} \,v^2 < 0    \, .
\end{equation}
Since $\mu_a^2< 0$, the extrema along the $h_{1,2} = 0$, $a \neq 0$ field direction of the scalar potential at zero temperature lie away from the origin of field space, at $\left\langle a \right\rangle = \pm v_S \equiv \pm \sqrt{\left|\mu_a^2 \right|/\lambda_a}$. For 
%
\begin{equation}
\label{saddle_vs_min}
(\lambda_{aH_1}v_S^2 +\mu_1^2)(\lambda_{aH_2}v_S^2 +\mu_2^2) > \mu_{12}^4 + \kappa^2 v_S^2,
\end{equation}
these are minima of the tree-level scalar potential (otherwise they are saddle points). We also require that the EW vacuum be the absolute vacuum of the theory, which yields
%
%
%
%
\begin{equation}
\label{vacuum_2}
- \mu^2_a  = \lambda_{\beta} \,v^2 - s_{\theta}^2 \,m_{a_2}^2 - c^2_{\theta} \, m_{a_1}^2  <  (m_h^2\, v^2)/ (2 \, v_S^2)\, .
\end{equation}
%
This yields a lower bound on $m_{a_1}$, given by 
\begin{equation}
\label{ma_min}
m_{a_1,\mathrm{min}}^2 = ([\lambda_{\beta}  - m_h^2/(2\, v_S^2)] c_{\theta}^2 \, v^2 - s_{\theta}^2 \, M^2)/(c_{\theta}^2 - s_{\theta}^2) \,
\end{equation}
and is automatically satisfied when the $\left\langle a \right\rangle$ extrema are saddle points.
%

%


\vspace{2mm}

\begin{figure}[t]
\begin{centering}
\includegraphics[width=0.48\textwidth]{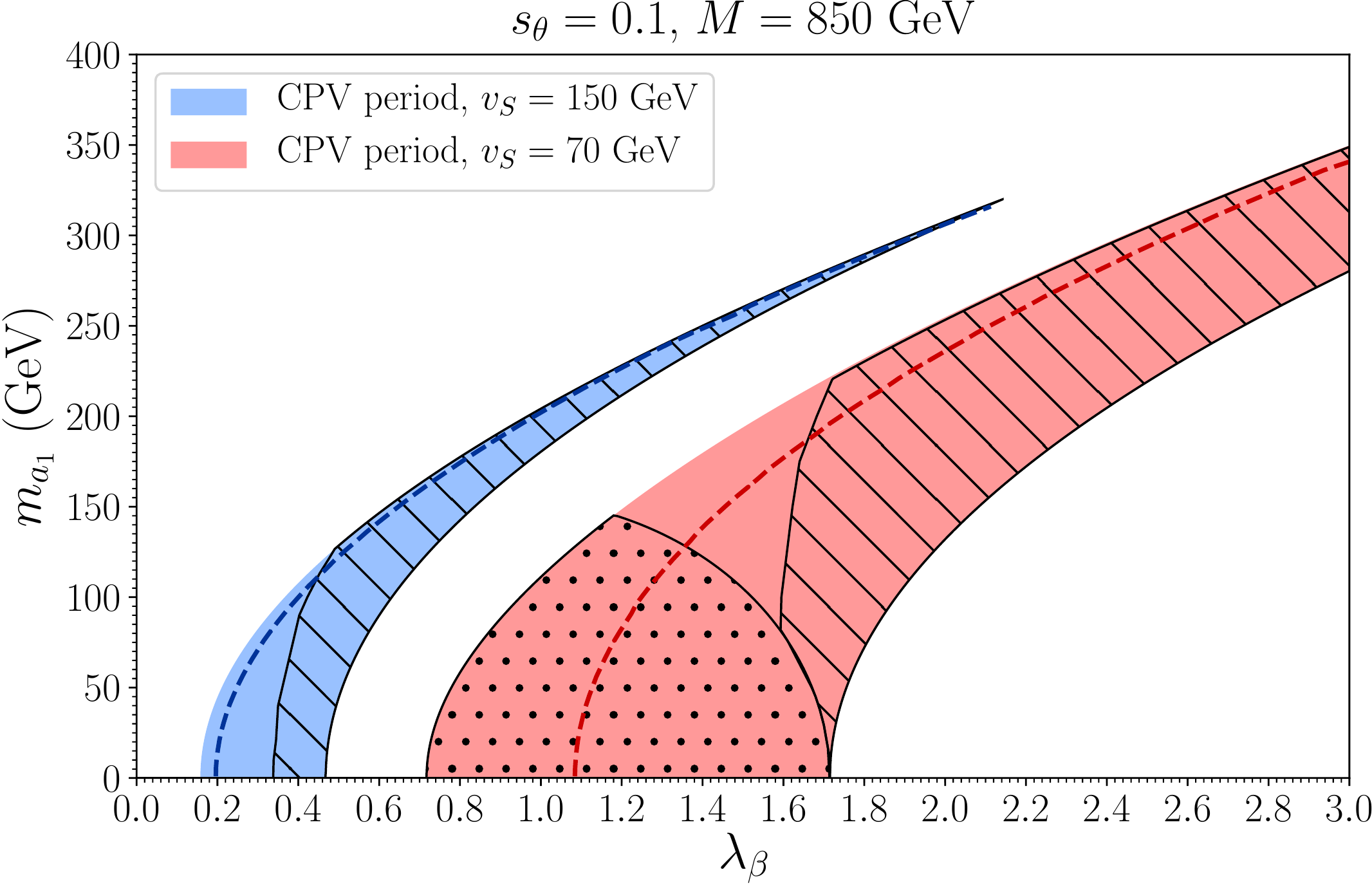}

\vspace{-2mm}

\caption{$m_{a_1}$ range leading to a spontaneous CP-breaking period prior to the EW phase transition, as a function of $\lambda_{\beta}$ 
for $s_\theta = 0.1$, $M = 850$ GeV and $v_S = 150$ GeV (blue), $v_S = 70$ GeV (red), respectively. The regions to the right of the corresponding dashed lines feature $\xi_c\gtrsim 1$ as needed for EW baryogenesis (see text for details). In the bar-hatched regions, the Universe is trapped in the CPV minimum (no $T_n$ exists). 
For the dot-hatched region, the CPV and EW extrema are saddle points at $T_c$ at $\mathcal{O}(T^2)$ (see footnote 2).}
\label{fig1:ma_max}
\par\end{centering}
\vspace{-2mm}

\end{figure}

The combined dynamics of the two Higgs doublets $H_{1,2}$ and the singlet field $a$ in the early Universe allows for a period of spontaneous CP breaking ending with the EW phase transition:
First, the singlet develops a non-zero vev $v_S(T)$ at a temperature $T_S$, with the EW symmetry remaining unbroken.~Then, EW symmetry breaking occurs at a lower temperature $T_h$, yielding a transition from the CPV minimum $(\left\langle h_1 \right\rangle, \left\langle h_2 \right\rangle, \left\langle a \right\rangle) = (0, 0, v_S(T))$ to the EW minimum $(v_1(T), v_2(T), 0)$, and restoring CP in the scalar sector.
This two-step symmetry breaking process can be described to a good approximation by 
keeping only the tree-level potential $V$ and the leading, $\mathcal{O}(T^2)$ thermal corrections in a high-temperature expansion of the 1-loop finite-temperature effective potential $V_{\mathrm{eff}}$ (see e.g.~\cite{Quiros:1999jp} for a review). The effective potential at $\mathcal{O}(T^2)$ has the advantage of manifestly avoiding issues related to the gauge-dependence of $V_{\mathrm{eff}}$, unlike the case with further contributions included~\cite{Patel:2011th}. In this letter, we keep our analysis at this order, leaving a study with the full 1-loop finite-temperature effective potential, including higher-order daisy contributions, for future work~\cite{HMN}. The $\mathcal{O}(T^2)$ thermal corrections are given by
\begin{equation}
V_T =  \frac{T^2}{24} \sum_b n_b \, M_b^2 + \frac{T^2}{48} \sum_f n_f M_f^2
\end{equation}
with $M_b(h_1, h_2, a)$ and $M_f(h_1, h_2, a)$ respectively the field-dependent masses for bosons and fermions, and $n_i$ the number of degrees of freedom of each species. Summing up the dominant contributions of the top quark, the $W$ and $Z$ gauge bosons, the 2HDM scalars $H^{\pm}$, $A_0$, $H_0$ and $h$, the Goldstone bosons $G^{\pm}$ and $G_0$, and the singlet $a$, $V_T$ reads (after dropping the field-independent contributions)
\begin{eqnarray}
\label{VT}
V_T &=& \frac{T^2}{24} \left[ \left(2 \lambda_3 + \lambda_4 + \frac{6 m_W^2 + 3 m^2_Z}{v^2}  \right) (h_1^2 + h^2_2) \right. \nonumber \\ 
  &+& \left.\left(3 \lambda_a + 4\lambda_{aH_1} + 4\lambda_{aH_2} \right) a^2 + \left(3 \lambda_1 + \lambda_{aH_1} \right) h_1^2 \right. \nonumber \\  
  &+& \left.\left(3 \lambda_2 + \lambda_{aH_2} + \frac{6 m_t^2 (1 + t_{\beta}^{-2})}{v^2}\right) h_2^2 \right] \, .
\end{eqnarray}
Then, as the temperature of the early Universe decreases during radiation domination, the temperature $T_S$ at which the singlet field direction gets destabilized (from the origin of field space) is given by 
$T_S^2 = 12 \left|\mu^2_a  \right|/(4\,\lambda_{aH_1} +4\,\lambda_{aH_2} + 3 \,\lambda_{a}) $.~The Higgs field directions get instead
%
%
destabilized at a temperature $T_h$ given by
$T_h^2 \simeq 6\, m_h^2 \, v^2 / \left(5 m_h^2 + \lambda_{\beta}\,v^2 + 6 m_W^2 + 3 m_Z^2 + 6 m_t^2 \right)$. 
%
%
Requiring $T_S > T_h$, for the spontaneous breaking of CP to occur prior to EW symmetry breaking, yields a lower bound on $\left|\mu^2_a  \right|$ which by virtue of~\eqref{mu2_negative} can be cast as an upper bound on $m_{a_1}$: 
%
%
%
%
\begin{equation}
\label{ma_max_T}
m_{a_1,\mathrm{max}}^2 = \frac{1}{c_{\theta}^2 - s_{\theta}^2} \left[c_{\theta}^2  
\,v^2 \, \lambda_{\beta} (1-F)  - s_{\theta}^2 \,M^2  \right] 
\end{equation}
with 
\begin{align}
\label{F_function}
F = \frac{(4\,\lambda_{aH_1} +4\,\lambda_{aH_2} + 3 \,\lambda_{a}) \, m_h^2}{2\left(5 m_h^2 + \lambda_{\beta} v^2 + 6 m_W^2 + 3 m_Z^2 + 6 m_t^2 \right)} \, .
\end{align}
The combination of Eqs.~\eqref{ma_min} and~\eqref{ma_max_T} then defines a specific region of the 2HDM $+$ $a$ parameter space where a period of spontaneous CP violation would take place in the early-Universe. This region is shown in Fig.~\ref{fig1:ma_max} in the ($\lambda_{\beta},\, m_{a_1}$) plane, for fixed $M$ and $s_\theta$, and two values of $v_S$ (we take here the singlet self-coupling $\lambda_a$ as a dependent parameter, subject to the perturbativity constraint $\lambda_a < 2\pi$). 



\vspace{1mm}

Within the parameter space 
defined by Eqs.~\eqref{ma_min} and~\eqref{ma_max_T}, there exists a {\it critical temperature} $T_c$ at which the CPV $(0, 0, v_S(T))$ and EW $(v_1(T), v_2(T), 0)$ extrema become degenerate. The interplay between tree-level and $\mathcal{O}(T^2)$ terms in the 1-loop finite-temperature effective potential generally gives rise to a potential barrier between the two extrema~\cite{Espinosa:2011ax} at $T_c$, which may result in a very strong first-order EW phase transition\footnote{For the dot-hatched parameter regions of Fig.~\ref{fig1:ma_max}, at $\mathcal{O}(T^2)$ in the effective potential the CPV and EW extrema are actually saddle points at $T_c$. The inclusion of further 1-loop contributions (e.g. Coleman-Weinberg) does however lift both saddle points by creating a potential barrier between them, still resulting in a first-order EW phase transition~\cite{HMN}. We thus retain those shaded regions as potentially baryogenesis-viable in the present analysis.}, as needed for EW baryogenesis.
The tunneling from the CPV to the EW minimum takes place at the {\it nucleation temperature} $T_{n} < T_c$, and we must ensure that tunneling does occur (i.e.~the Universe does not stay trapped in the singlet vacuum, see e.g.~\cite{Baum:2020vfl,Biekotter:2021ysx} for recent discussions) for EW symmetry breaking to occur and our scenario to be physically viable.
To verify this is the case, we implement the 2HDM + $a$ in the numerical code~\texttt{CosmoTransitions}~\cite{Wainwright:2011kj}. The unphysical region for which the Universe becomes trapped in the CPV minimum (and no $T_n$ exists) is shown as bar-hatched in Fig.~\ref{fig1:ma_max}.

The strength of the transition (which quantifies the departure from thermal equilibrium) can be characterized by the parameters $\xi_{c(n)}={v_{c(n)}}/{T_{c(n)}}$, 
with $v_{c(n)}$ the value of the EW vev at $T_{c(n)}$. Successful baryogenesis requires $\xi_n \gtrsim 1$, to avoid baryon number washout. Except for the strongest transitions, $\xi_c$ gives a good estimate of $\xi_n$, and we use the former in our discussion below (leaving a more detailed analysis for the future). Note that this approximation underestimates the relevant strength of the phase transition and the corresponding baryon asymmetry (see the discussion around Eq.~\eqref{baryo_approx}).

\vspace{2mm}

\noindent \textbf{CP violation.}~CP violation in the scalar sector is encoded in the phase of quantities that are rephasing invariant under $U(1)$ transformations of the Higgs doublets, e.g., $\lambda_5^* (\mu_{12}^2)^2$ in the 2HDM~\cite{Inoue:2014nva}. For $\lambda_5 = 0$ (following from our $m_{H_0}^2 \simeq m_{A_0}^2 \simeq m_{H^{\pm}}^2 = M^2$ simplifying assumption) there is thus no CP violation in the 2HDM scalar sector. In contrast, in the 2HDM $+$ $a$, the presence of the 
$a H^\dagger_1 H_2$ term in $V$ yields the additional physical CP-violating phase $\delta_{\kappa} = {\rm Arg} [\kappa^* \mu_{12}^2]$.~For $\mu_{12}^2,\, \kappa \in \mathbb{R}$ (so that $\delta_{\kappa} = 0$) the scalar sector of the 2HDM $+$ $a$ conserves CP in the EW minimum and there are no contributions to EDMs beyond the SM. Yet, even in this case a non-zero singlet vev $\left\langle a \right\rangle = v_S(T)$ generates transient CP violation via a complex 
squared-mass coefficient $\mu^{2}_{12}(T)$ for the $H_1^\dagger H_2$ term in~\eqref{2HDM_potential},   
\begin{equation}
\mu^{2}_{12}(T) = \mu_{12}^2 - i\, \kappa\, v_S(T)   \, .
\end{equation}
The CP-violating phase $\delta_S = {\rm Arg} [ \mu_{12}^2(T)^* \mu_{12}^2 ]$  is physical, signaling a non-removable phase difference between vacua (here, the CPV minimum and the EW minimum) with different values of $v_S(T)$. 

\begin{figure}[t]
\begin{centering}
\includegraphics[width=0.5\textwidth]{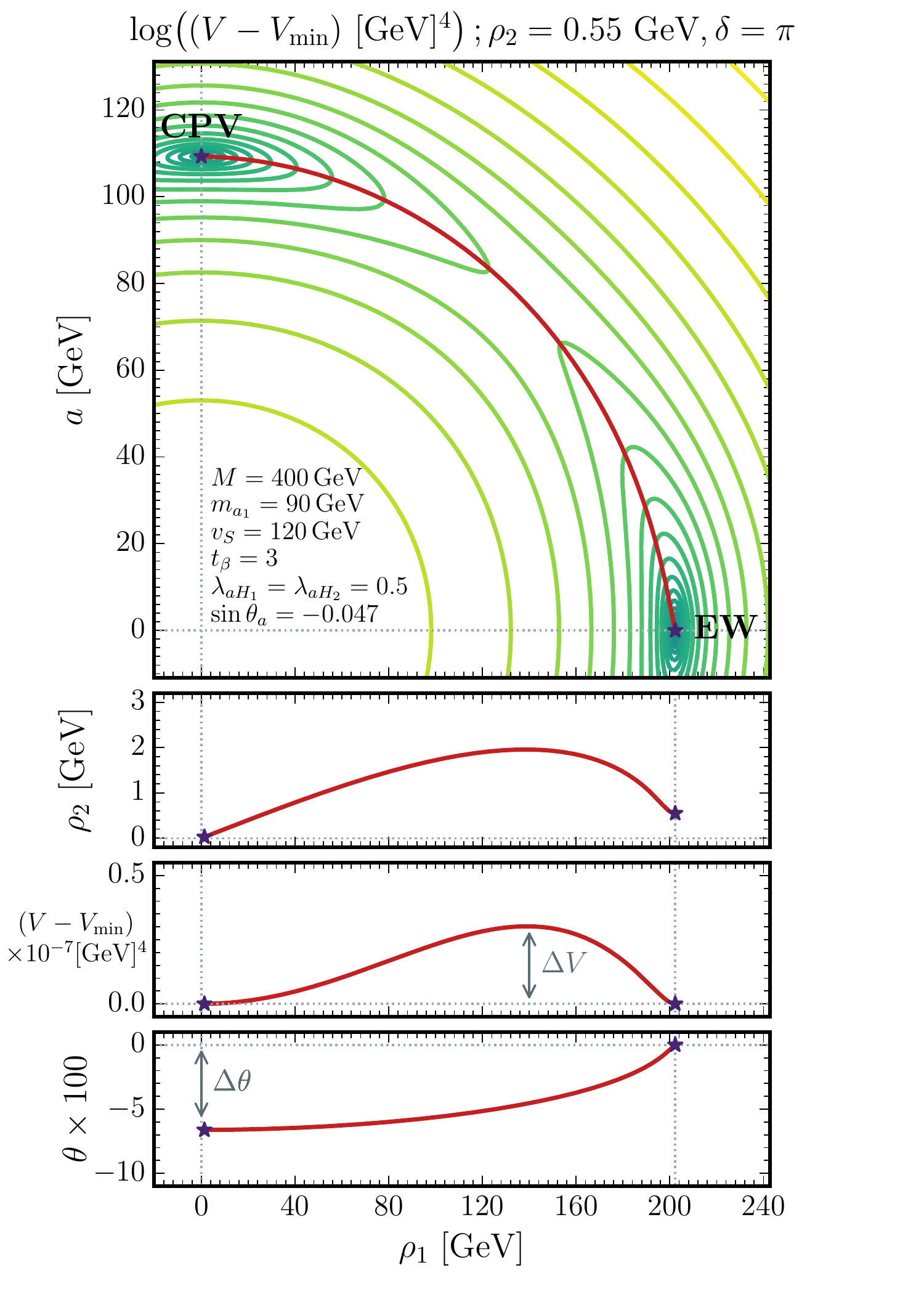}

\vspace{-2mm}

\caption{Visualisation of the potential in $(\rho_1,a)$ space, for a benchmark point in our simplified scenario, with $M = 400$ GeV, $m_{a_1}= 90$ GeV, $v_S= 120$ GeV, $\lambda_{aH_1}=\lambda_{aH_2}=0.5$, $t_\beta=3$ and $\sin\theta =-0.047$.
The potential is evaluated at $T = T_c$, fixing the values of $\rho_2$ and $\delta$ to their minima (0.55 GeV and $\pi$, respectively) in the EW phase. The path in field space that minimises the effective potential is shown in red. The insets show the variation along this trajectory of the second Higgs field, $\rho_2$, the value of the potential and the relative phase between the two doublet fields in the $\mathbb{Z}_2$ basis, $\theta$ (See text for details).
}
\label{fig2:potential_contours}
\par\end{centering}
\vspace{-2mm}

\end{figure}

Expressing the neutral components of the Higgs doublets in radial form,
\begin{align}
        h_1+i\, \zeta_1 = \eta_1 e^{i\theta_1},\quad h_2+i\, \zeta_2= \eta_2 e^{i\theta_2},
\end{align}
only the combination $\theta=(\theta_1-\theta_2)$ appears in the scalar potential $V$.~The variation of $\theta$ during the EW phase transition due to the trajectory in field space between the CPV and EW minima\footnote{Strictly speaking, $\theta$ is not defined at the CPV minimum, since both Higgs vevs are zero. However, it is well-defined by its limiting value, as the field trajectory tends to the CPV minimum.} leads to the pseudoscalar component of the two Higgs doublets acquiring a transient non-zero value, before settling in the usual EW minimum at zero-temperature. The net change in $\theta$ across the EW phase transition precisely corresponds to $\delta_S$.
%
%
It is convenient to rotate the pair of Higgs doublets into the so-called ``Higgs-basis''~\cite{Gunion:2002zf}, where only one of the doublets acquires a vev at zero-temperature, in contrast to the ``$\mathbb{Z}_2$-basis'', where the softly-broken discrete symmetry is manifest but both doublets participate in EWSB. The rotation is achieved by the angle $\beta$ and the analogous non-linear representation defines (recall that we are working in the 2HDM alignment limit):
\begin{align}
        h+i\, g = \rho_1 e^{i\delta_1},\quad H_0+i\, A_0= \rho_2 e^{i\delta_2},
\end{align}
where the imaginary components of the doublets now correspond to the would-be neutral Goldstone mode and the 2HDM pseudoscalar, respectively. The physical phase in this case is given by $\delta=(\delta_1-\delta_2)$. 
The (4-dimensional) field trajectory between the CPV and EW  minima is formally determined by finding the `bounce' solution that extremizes the Euclidean action, yet it is well-approximated by the field trajectory which minimises the effective potential.
Here, we obtain that trajectory numerically by tracing along the straight line between the two minima in $(\rho_1,a)$ space, and minimizing along the three remaining field directions.
In Fig.~\ref{fig2:potential_contours} (top) we show iso-contours of the effective potential at $T = T_c$ in the $(\rho_1,a)$ plane 
for a chosen benchmark. The variation of several quantities along the field space trajectory as a function of $\rho_1$ are shown in the subfigures. The top subfigure shows the variation of the modulus of the second doublet, $\rho_2$. The middle subfigure shows the value of the potential, and hence the shape of the potential barrier between the two minima. The bottom subfigure shows the variation of $\theta$ and hence the amount of transient CP-violation generated during the phase transition. This is characterised by the change in $\theta$, $\Delta\theta = \delta_S$, between the CPV and EW minima.

%
%
%
%
%

\vspace{2mm}

%
%
%
\noindent \textbf{Baryogenesis.}~Electroweak baryogenesis is driven by CP-violating interactions of the SM fermions with the expanding Higgs bubble walls (similarly to what occurs in baryogenesis within the 2HDM~\cite{Fromme:2006cm,Dorsch:2016nrg}). This generates chiral densities of quarks which diffuse into the symmetric phase in front of the bubbles. Electroweak sphalerons then produce a net baryon number. In the setup considered here, the top quark is the relevant fermion, and we use the formalism presented in~\cite{Fromme:2006wx}. It used to be commonly assumed that successful baryogenesis would require sub-sonic bubble expansion to allow for efficient diffusion of chiral densities into the symmetric phase (as e.g. done in~\cite{Fromme:2006wx}), but recently it was shown that this assumption is too restrictive~\cite{Cline:2020jre,Dorsch:2021ubz}, and baryogenesis is also possible for larger values of the bubble wall velocity $v_W$. 
Rather than solving the transport equations of~\cite{Fromme:2006wx} explicitly, we use the analytic approximation 
\begin{equation}
\label{baryo_approx}
\frac{\eta}{10^{-11}} 
\sim 6\times 10^2  \, \frac{\rm{sin}(\delta_{t})\,\xi_c^2}{L_{W} T_{c}}
\end{equation}
where $\eta$ is the baryon-to-entropy ratio (our measure of the generated BAU), $L_W$ is the thickness of the bubble wall 
and $\delta_t$ is the variation of the complex phase of the top-quark mass along the bubble wall, which drives the baryogenesis process and is related to $\delta_S$ by $\delta_t = \delta_S/(1 + t_\beta^2)$.
Comparing with~\cite{Fromme:2006wx}, \eqref{baryo_approx} reproduces the full results to about a factor of 2 for $v_W\lesssim 0.2$, which is sufficient accuracy for our application. In the following we thus allow $\eta \in [\eta_{\rm OBS}/2,\,2\, \eta_{\rm OBS}]$, with $\eta_{\rm OBS}$ the observed BAU $\eta_{\rm OBS} = 8.7 \times 10^{-11}$~\cite{Planck:2015fie}.
%
The results of~\cite{Fromme:2006wx} are based on a derivative expansion of the transport equations so require a sufficiently thick bubble wall, $L_W\, T_c \gtrsim 2$, which we find to be generically satisfied across the parameter space we study. %

On general grounds, from the results of~\cite{Cline:2020jre,Dorsch:2021ubz} one expects a mild suppression of the BAU ($\sim 50\%$) compared to~\eqref{baryo_approx} for $0.2 \lesssim v_W \lesssim 0.5$, while for even larger wall velocities the suppression may become more severe. A detailed assessment of the impact of $v_W$ on our BAU estimates is however beyond the scope of this work, and we use the approximation~\eqref{baryo_approx} in its present form.

\begin{figure}[t]
\begin{centering}
\includegraphics[width=0.49\textwidth]{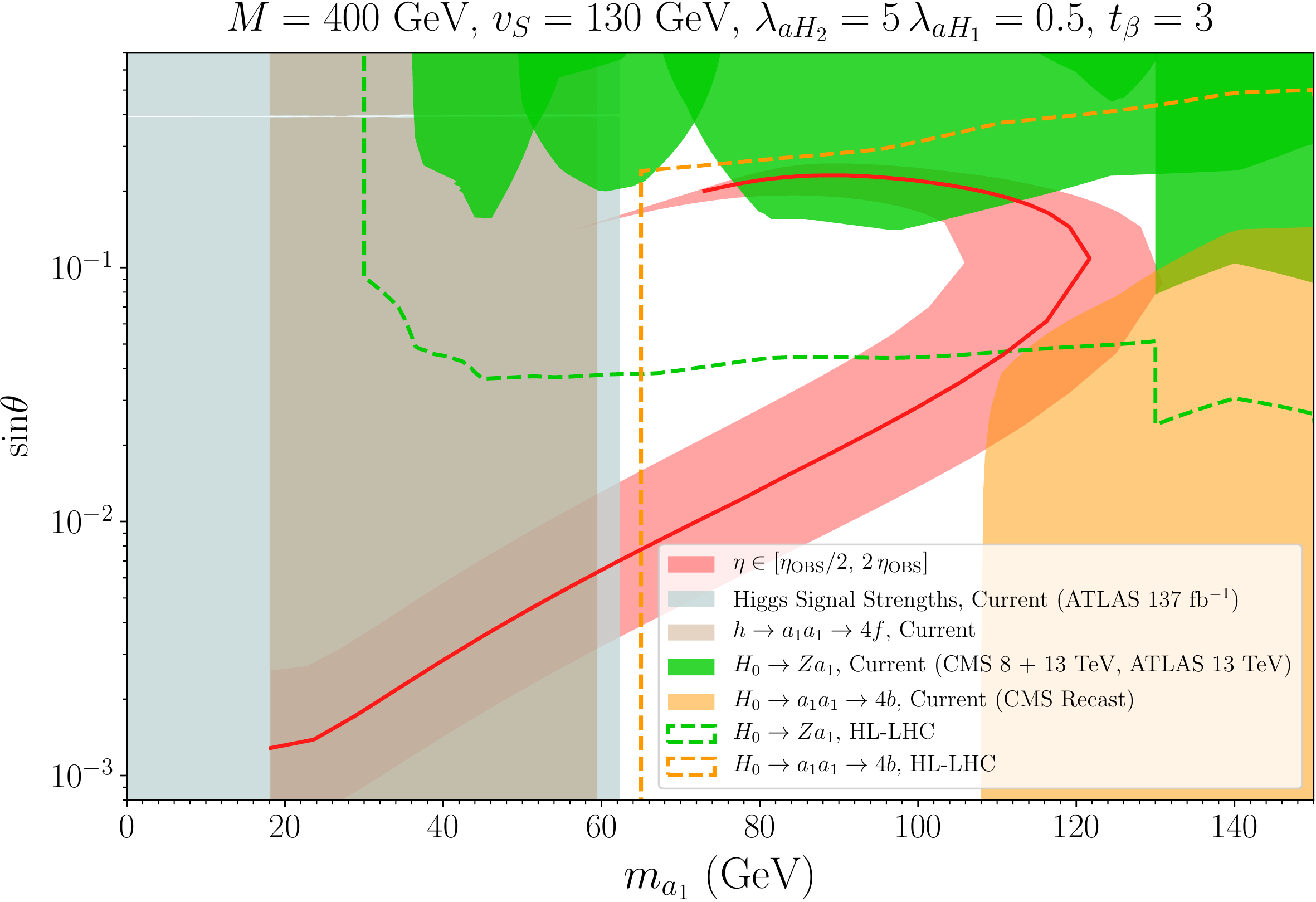}

\vspace{2mm}

\includegraphics[width=0.49\textwidth]{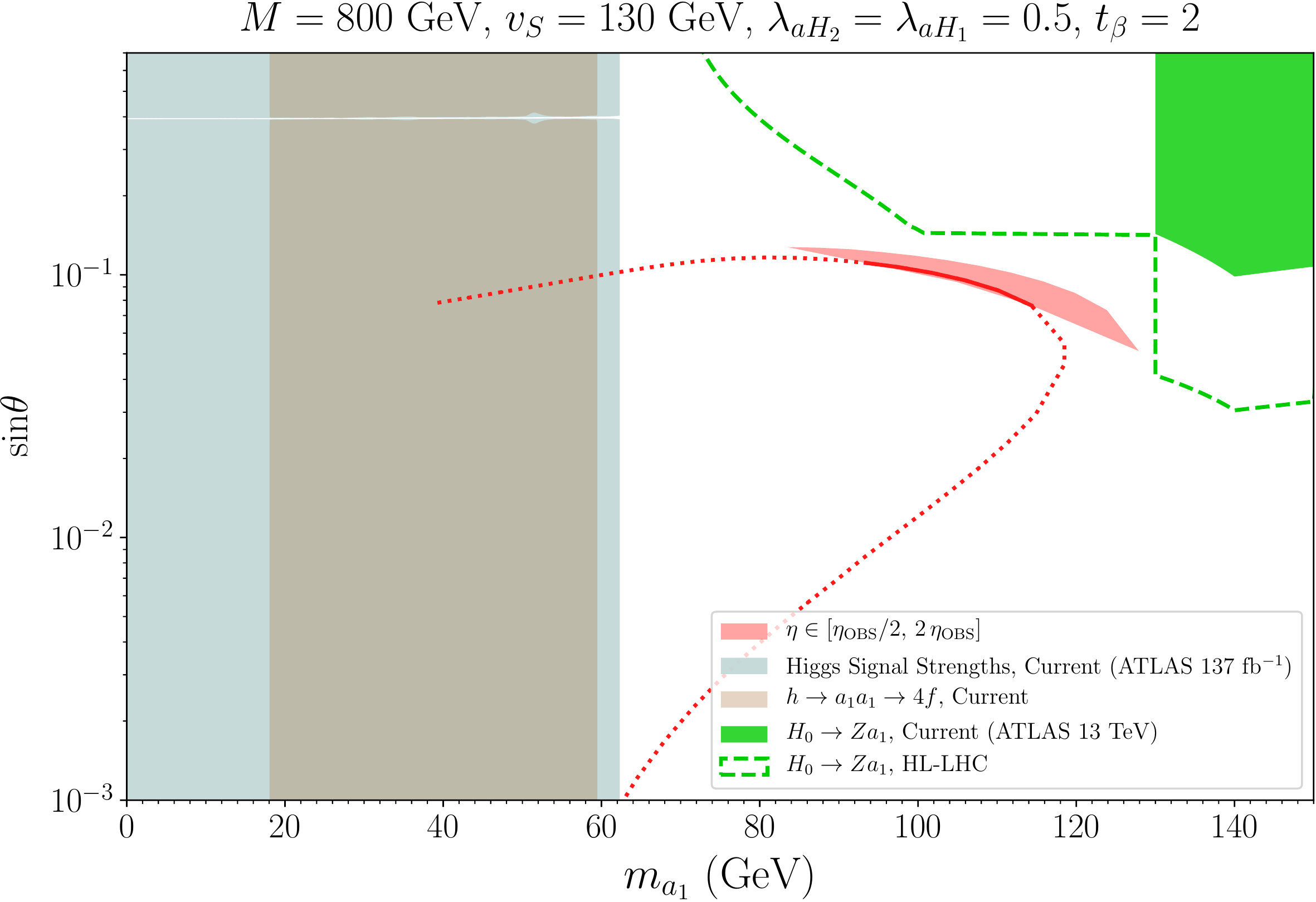}

\caption{($m_{a_1}$, $s_{\theta}$) plane for the benchmarks $M = 400$ GeV, $v_S = 130$ GeV, $t_\beta = 3$, $\lambda_{aH_2} = 5\, \lambda_{aH_1} = 0.5$ in Type-I 2HDM (top panel) and $M = 800$ GeV, $v_S = 130$ GeV, $t_\beta = 2$, $\lambda_{aH_2} = \lambda_{aH_1} = 0.5$ in Type-II 2HDM (bottom panel). The red band corresponds to $\eta \in [\eta_{\rm OBS}/2,\,2\, \eta_{\rm OBS}]$, with $\eta = \eta_{\rm OBS}$ shown as a solid-red line. Parameter values that would yield $\eta = \eta_{\rm OBS}$ (based on an analysis at $T_c$) but for which no $T_n$ exists (thus being unphysical) are shown as a dotted-red line. Solid coloured regions (green, grey, brown, yellow) correspond to present experimentally excluded regions by LHC searches, while the dashed-coloured lines show the future HL-LHC 95 \% C.L. exclusion sensitivity in each case.}
\label{fig3:BAU}
\par\end{centering}
\vspace{-2mm}

\end{figure}

In Fig.~\ref{fig3:BAU} we show in the ($m_{a_1}$, $s_{\theta}$) plane two representative 2HDM$+ a$ baryogenesis benchmarks, given by $M = 400$ GeV, $v_S = 130$ GeV, $t_\beta = 3$, $\lambda_{aH_2} = 5\, \lambda_{aH_1} = 0.5$ (in Type-I 2HDM, see~\cite{Branco:2011iw}), and $M = 800$ GeV, $v_S = 130$ GeV, $t_\beta = 2$, $\lambda_{aH_2} = \lambda_{aH_1} = 0.5$ (in Type-II 2HDM), depicted respectively in Fig.~\ref{fig3:BAU} - top and Fig.~\ref{fig3:BAU} - bottom. In each case, the region yielding $\eta \in [\eta_{\rm OBS}/2,\,2\, \eta_{\rm OBS}]$ is shown as a red band. The benchmarks in Fig.~\ref{fig3:BAU} - top never feature a potential barrier between the EW minimum and the CPV extrema at $T = 0$ (since Eq.~\eqref{saddle_vs_min} is never satisfied), and thus $T_n$ is guaranteed to exist within our approximations. In contrast, for the benchmarks in Fig.~\ref{fig3:BAU} - bottom, there is a set of $m_{a_1}$ vs $s_{\theta}$ values that would yield $\eta = \eta_{\rm OBS}$ based on our analysis at $T_c$ but for which no $T_n$ exists (and is thus unphysical). These are depicted in Fig.~\ref{fig3:BAU} as a dotted-red line. 

At this point, we stress that, in the EW symmetric phase, the scalar potential $V_{a}$ has a $\mathbb{Z}_2$ symmetry under which $a \to -a$, resulting in equal transition probability to $\pm v_S(T)$ when CP is spontaneously broken. This jeopardizes successful baryogenesis, as the asymmetry generated in regions with $+v_S(T)$ would be balanced by that generated in regions with $-v_S(T)$, since $\delta_S$ is of opposite sign. An explicit $\mathbb{Z}_2$-breaking term in $V_a$ given by, e.g., $\mu_3\, a^3$ solves this issue by introducing a bias between the two singlet vacua, given approximately by (see the Appendix) $\Delta V \simeq \mu_3 \, v_S (T)^3$.~The bias needed for the regions in the deeper minimum to completely dominate when the EW phase transition takes place is tiny, $\Delta V/ T_n^4 \gg 10^{-16}$ (see~\cite{McDonald:1993ey,Espinosa:2011eu}).
From the properties of $a$, it is clear that CP breaking and $\mathbb{Z}_2$ breaking are connected: an explicit breaking of CP in $V$ (e.g. through a complex $\mu^2_{12}$ term) would radiatively generate an $a^3$ term. 
Thus, it is plausible that CP violation in the SM quark sector, leaking to the 2HDM scalar sector (as recently discussed in~\cite{Fontes:2021znm}) can generate a sufficient bias, as we discuss in the Appendix. In any case, a very small amount of explicit CP breaking in the 2HDM scalar sector, far below foreseen EDM experimental sensitivity (see e.g.~\cite{Inoue:2014nva,Dorsch:2016nrg}) is enough to provide the required bias for successful baryogenesis, and CP is 
approximately conserved at present in the scalar sector.

\vspace{2mm}

\noindent \textbf{Phenomenological implications.}~The landmark phenomenological signatures of the scenario investigated in this work are the existence of a light pseudoscalar $a_1$ (with $m_{a_1} < v$) accompanied by a (2HDM-like) set of spin-$0$ states around/below the TeV scale (recall that $m_{H_0} \simeq M$, and Eq.~\eqref{ma_max_T} imposes $M < v \sqrt{\lambda_{\beta}} /t_{\theta}$ to obtain $m_{a_1, \rm{max}} > 0$), with cascade decays into $a_1$, e.g. $H_0 \to Z a_1$, $H_0 \to a_1 a_1$, $a_2 \to h a_1$ and $H^{\pm} \to W^{\pm} a_1$. In Fig.~\ref{fig3:BAU} we show the current experimental limits and future sensitivity to the two representative 2HDM$+ a$ baryogenesis benchmarks discussed above, in the $m_{a_1}$ vs $s_{\theta}$ plane. 
%
%
%
%
The mass region $m_{a_1} < m_h/2$ is very strongly constrained by both LHC measurements of the 125 GeV Higgs signal strengths~\cite{ATLAS:2021vrm} and direct searches for exotic Higgs boson decays $h \to a_1 a_1 \to 4 f$ ($f$ being SM fermions), e.g.~in the $b \bar{b} \tau \tau$ and $b \bar{b} \mu \mu$ final states (see~\cite{Cepeda:2021rql} for an  up-to-date review of these searches). These probe the branching fraction BR$(h \to a_1 a_1)$ (proportional to $\lambda_{\beta}^2 \, c_{\theta}^4$ and whose explicit expression in terms of the 2HDM$+a$ parameters can be found in~\cite{Bauer:2017ota}), and Fig.~\ref{fig3:BAU} shows the corresponding present 95\% C.L. limits.

\vspace{1mm}

The specific interplay among different experimental probes of the light pseudoscalar $a_1$ in LHC cascade decays (when available by phase-space) heavily depends on the value of $s_{\theta}$. The decay widths for $H_0 \to Z a_1$, $a_2 \to h a_1$ and $H^{\pm} \to W^{\pm} a_1$ are proportional to $s_{\theta}^2$, and thus such channels may become suppressed for $s_{\theta} \ll 1$. Yet, in this limit the decay width for $H_0 \to a_1 a_1$ scales as $(\lambda_{aH_1} - \lambda_{aH_2})^2 \, c_{\theta}^4$~\cite{Bauer:2017ota} and could thus lead to a large BR for $\lambda_{aH_1} \neq \lambda_{aH_2}$. 
%
%
%
%
%
In particular, for $|\lambda_{aH_1} - \lambda_{aH_2}| \gtrsim 0.1$, $H_0 \to a_1 a_1$ can be the dominant decay channel of $H_0$, and would constitute an important probe of the $s_{\theta} \ll 1$ baryogenesis regime. Ref~\cite{Barducci:2019xkq} has recently performed an LHC sensitivity study of the channel $p p \to H_0 \to a_1 a_1 \to b\bar{b} b\bar{b}$ by recasting the latest di-Higgs CMS search in the $4b$ final state~\cite{CMS:2018qmt}.
We here use \texttt{SusHi}~\cite{Harlander:2012pb} to obtain the NNLO production cross section for $H_0$ and show in Fig.~\ref{fig3:BAU} the present 95\% C.L. limit from searches for the $p p \to H_0 \to Z a_1 \to \ell \ell \bar{b} b$ channel by CMS~\cite{CMS:2016xnc,CMS:2019wml} and ATLAS~\cite{ATLAS:2018oht}, and searches for $p p \to H_0 \to a_1 a_1 \to b\bar{b} b\bar{b}$ (from the recast performed in~\cite{Barducci:2019xkq}). We also show the expected 95\% C.L. sensitivity of all these searches at the HL-LHC as dashed lines. For the benchmark in Fig.~\ref{fig3:BAU} - top, the combination of these searches with Higgs signal strength measurements will be able to probe the whole region of viable baryogenesis. This is however not entirely so for the benchmark in Fig.~\ref{fig3:BAU} - bottom, for which  $\lambda_{aH_1} = \lambda_{aH_2}$ and thus $H_0 \to a_1 a_1$ searches do not provide sensitivity. We also find that other LHC searches for $a_1$ (e.g. direct production in gluon-fusion followed by the $a_1$ decay into di-photon or di-tau final states) and for $H_0$ (e.g. via $H_0 \to \tau \tau$ decays)
do not provide meaningful constraints to the benchmarks considered in Fig.~\ref{fig3:BAU}.
%
%
%
%
%
%

Finally, the viable baryogenesis parameter space is also constrained by flavour observables, particularly from rare $B$-meson decays: Existing constraints from $b \to s \gamma$ decays set a 95\% C.L. limit $m_{H^{\pm}} > 790$ GeV for Type-II 2HDM~\cite{Atkinson:2021eox} (the specific value of this limit being currently under debate~\cite{Hermann:2012fc,Misiak:2020vlo}), taken into account in our analysis. The existence of a light pseudoscalar state $a_1$ coupling to SM fermions could also be probed by its contributions to the decay $B_s\to\mu^+\mu^-$~\cite{Skiba:1992mg,Logan:2000iv}, but only for Type-II 2HDM and $t_\beta \gg 1$, which would however suppress the generated BAU.

\vspace{2mm}

\noindent \textbf{Comment on the DM connection.}~The 2HDM + $a$ model is a well-motivated DM portal, with the simple addition of a singlet Dirac fermion $\chi$ (the DM candidate) coupled to $a$ through a Yukawa term $y_\chi \, a \bar{\chi} \gamma_5 \chi$. Among its virtues, it naturally avoids spin-independent DM direct detection experimental limits (since $a$ is a pseudoscalar mediator, see e.g.~\cite{DeSimone:2016fbz}), and  could explain the possible galactic center $\gamma$-ray excess~\cite{Ipek:2014gua,Hooper:2010mq,Hooper:2011ti}. Remarkably, there exists an interplay between baryogenesis and DM in this scenario: The coupling $y_\chi$ contributes to the effective potential, and its effect is included via $4\,\lambda_{aH_1} +4\,\lambda_{aH_2} + 3 \,\lambda_{a} \to 4\,\lambda_{aH_1} +4\,\lambda_{aH_2} + 3 \,\lambda_{a} + 2\, y_\chi^2$ in~\eqref{F_function}. Thus, it leads to a decrease in $T_S$ and the corresponding shrinking of the parameter space region where spontaneous CP violation prior to the EW phase transition occurs. At the same time, the 
early Universe DM annihilation cross section for $\chi\bar\chi \to p_{\rm SM} \, p_{\rm SM}$ (with $p_{\rm SM}$ generic SM particles) processes leading to the observed DM relic density via thermal freeze-out scales as $\langle \sigma \mathrm{v}\rangle \propto y_\chi^2\, s_{\theta}^2 \,m_\chi^2 / m_{a_1}^4$ (for $m_{a_1} \ll  m_{a_2}$). With the rest of model parameters fixed, the value of $y_\chi$ yielding the observed DM relic density scales as $s_{\theta}^{-2}$, leading to a minimum value of $s_{\theta}$ compatible with the DM relic density and yielding spontaneous CP violation prior to the EW phase transition (i.e. with $T_S > T_h$).
We will explore this interplay in detail and the possibility of achieving baryogenesis and the correct relic DM abundance in this scenario in an upcoming work~\cite{HMN}.

\vspace{2mm}

\noindent \textbf{Conclusions.}~We have proposed a variant of electroweak baryogenesis characterized by the spontaneous breaking of CP in the early Universe, driven by the vev of a CP-odd scalar $a$. Working in a specific realization of such a setup, given by a 2HDM + $a$ extended Higgs sector, we have shown that this CP-breaking period, followed by a strongly first-order electroweak phase transition, is able to generate the observed baryon asymmetry of the Universe in sizable regions of the parameter space of the model, while naturally avoiding the stringent EDM experimental constraints on BSM CP-violating sources. The existence of such a CP-breaking period in the early Universe generally demands the existence of a singlet-like light CP-odd scalar (necessarily coupled to the SM fermions (via its mixing with the Higgs doublets) to ensure viable baryogenesis. This light pseudo-scalar can be searched for at the LHC, dominantly in cascade decays of the heavier 2HDM scalars (and also possibly via low-energy flavour measurements), yielding a powerful probe of our proposed scenario. Finally, we stress that the electroweak phase transition in this setup is rather strong in general, and would possibly lead to a stochastic gravitational wave signal in the observable range of the future LISA observatory~\cite{Caprini:2019egz}, a study we intend to carry out in the future. 

\vspace{2mm}

\begin{center}
\textbf{APPENDIX: Explicit CP violation from the SM}
\end{center}

\begin{figure}[t]
\begin{centering}
\includegraphics[width=0.34\textwidth]{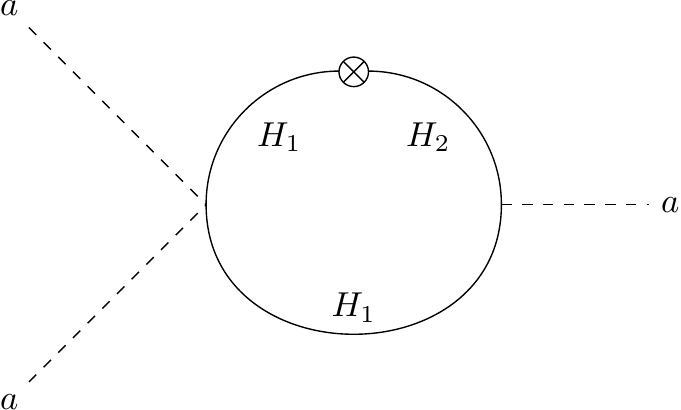} 

\caption{1-loop contribution to the CP-breaking term $a^3$, proportional to $\lambda_{a H_1}$. The $\otimes$ symbol represents an insertion of the $\mu^2_{12}$ mass term from $V_{{\rm 2HDM}}$.}
\label{fig1:Feynman_S3}
\par\end{centering}

\vspace{-2mm}

\end{figure}

As outlined in the main text, the scalar potential $V_{a}$ presents an unbroken $\mathbb{Z}_2$ symmetry under which $a \to -a$, (with either $H_1$ or $H_2$ being also odd under the $\mathbb{Z}_2$). When CP is spontaneously broken, there will not be a bias between transitions to $-v_S(T)$ and $+v_S(T)$, which would hinder viable baryogenesis: the volume of the regions with $+v_S(T)$ and $-v_S(T)$ would be essentially equal, and the baryon asymmetry generated in the regions with $+v_S(T)$ would be cancelled out by that generated in the regions with $-v_S(T)$, since $\delta_S \propto {\rm sign}[\pm v_S(T)]$. This issue is solved if a small explicit $\mathbb{Z}_2$ breaking term in $V_a$ exists, e.g. $(\mu_3/3) a^3$. Such a term would introduce a bias between the two singlet vacua, given 
by 
\begin{equation}
\label{vacuum_S_bias}
\Delta V =  \frac{\mu_3(\mu_3^2 + \lambda_a \mu_a^2)\sqrt{\mu_3^2 + 4 \lambda_a \mu_a^2}}{6 \lambda_a^3}
\end{equation}
The amount of bias $\Delta V$ needed for the (deeper) minimum to be the only vacuum remaining when the EW phase transition takes place is tiny, $\Delta V/ T^4 \gg 10^{-16}$ (see~\cite{McDonald:1993ey,Espinosa:2011eu}).
Then, assuming $\mu_3^2 \ll \lambda_a \mu_a^2$ near the onset of the EW phase transition, we would require 
\begin{equation}
\label{vacuum_Universe}
\frac{\mu_3 \, v_S (T)^3}{T^4} \gg 10^{-16}
\end{equation}
At the same time, an $a^3$ term in $V_a$ would lead to an explicit breaking of CP in the scalar potential, and so the radiative generation of a non-zero $\mu_3$ is (only) possible via an explicit breaking of CP in $V_{\mathrm{2HDM}} + V_a$, e.g. through a complex $\mu^2_{12}$ term (when $\kappa$ is real). 
This term would generate $\mu_3$ at 1-loop (see Figure~\ref{fig1:Feynman_S3}), of order 
%
%
\begin{eqnarray}
\label{vacuum_S3}
\mu_3 & \sim & \int \frac{d^4 p}{(4\,\pi)^2}\,  \frac{{\rm max}(\lambda_{aH_1},\, \lambda_{aH_2})\,  \kappa \,\left|\mu_{12}^2\right| \, \mathrm{sin}\, \delta_{12}}{(p^2 - M^2)^3} \nonumber \\ 
&\sim & \frac{{\rm max}(\lambda_{aH_1},\, \lambda_{aH_2})\,\kappa \,\left|\mu_{12}^2\right| \, \mathrm{sin} \, \delta_{12}}{16\,\pi^2 M^2}\, ,
\end{eqnarray}

\noindent with $\delta_{12}$ the phase of the $\mu^2_{12}$ term for real $\kappa$. For ${\rm max}(\lambda_{aH_1},\, \lambda_{aH_2}) \sim \mathcal{O}(1)$, sin$\,\theta \ll 1$ (such that $\kappa \sim {\rm sin}\, \theta \times M^2/v$), $\left|\mu_{12}^2\right| \sim M^2$ and $v_S(T) \sim T$, the condition~\eqref{vacuum_Universe} for viable baryogenesis translates into 
\begin{equation}
\label{minimum_CP}
\mathrm{sin}\, \delta_{12} \gg 10^{-16} \times \frac{(16\, \pi^2)\, v\, T}{M^2} \sim 10^{-15}
\end{equation}
where a mild hierarchy between $M$ and both $v$ and $T_n$ has been assumed in the last step of~\eqref{minimum_CP}.  

In this respect, it has been recently pointed out~\cite{Fontes:2021znm} that, given that CP is violated in the SM by the fermion sector, this will inevitably leak to the scalar sector, and renormalizability of the 2HDM at the 3-loop level does seem to require a complex $\mu_{12}^2$, i.e.~a non-zero value of $\delta_{12}$. Assuming a 3-loop suppression combined with the SM Jarlskog invariant suppression for the generated $\delta_{12}$ yields $\mathrm{sin}\, \delta_{12} \sim 10^{-12}$, very far below current and foreseen EDM experimental sensitivity (see e.g.~\cite{Inoue:2014nva,Dorsch:2016nrg}) but still above what is needed to satisfy~\eqref{minimum_CP}. 
Thus, it is plausible that CP violation in the SM quark sector, leaking to the 2HDM scalar sector (as recently discussed in~\cite{Fontes:2021znm}) can actually generate the required bias in $V_a$ for successful baryogenesis. 



\vspace{3mm}

\begin{center}
\textbf{Acknowledgements} 
\end{center}

\vspace{-1mm}

\begin{acknowledgements}

Feynman diagrams were drawn using {\sc TikZ-Feynman}~\cite{Ellis:2016jkw}. S.J.H. is supported by the Science Technology and Facilities Council (STFC) under Consolidated Grant ST/T00102X/1.
K.M. is supported by the UK STFC via grant
ST/T000759/1.
The work of J.M.N. is supported by the Ram\'on y Cajal Fellowship contract RYC-2017-22986, and by grant PGC2018-096646-A-I00 from the Spanish Proyectos de I+D de Generaci\'on de Conocimiento.
J.M.N. also acknowledges support from the European Union's Horizon 2020 research and innovation programme under the Marie Sklodowska-Curie grant agreement 860881 (ITN HIDDeN), as well as from  
the grant IFT Centro de Excelencia Severo Ochoa
CEX2020-001007-S funded by MCIN/AEI/10.13039/501100011033.

\end{acknowledgements}

\bibliography{CPV}

\end{document}